\documentclass[fleqn,10pt]{wlscirep}
\usepackage[pdftex]{dropping}

\title{Cargo binding promotes KDEL receptor clustering at the 
mammalian cell surface}

\author[1,*]{Bj\"orn Becker}
\author[2,*]{M.\ Reza Shaebani}
\author[1]{Domenik Rammo}
\author[1]{Tobias Bubel}
\author[2]{Ludger Santen}
\author[1,+]{Manfred J. Schmitt}
\affil[1]{Molecular and Cell Biology, Department of Biosciences 
and Center of Human and Molecular Biology (ZHMB), Saarland 
University, D-66041 Saarbr\"ucken, Germany}
\affil[2]{Department of Theoretical Physics, Saarland University, 
D-66041 Saarbr\"ucken, Germany}
\affil[*]{B.B. and M.R.S. contributed equally to this work.}
\affil[+]{Correspondence should be addressed to mjs@microbiol.uni-sb.de}

\begin{abstract}
Transmembrane receptor clustering is a ubiquitous phenomenon in 
pro- and eukaryotic cells to physically sense receptor/ligand 
interactions and subsequently translate an exogenous signal into 
a cellular response. Despite that receptor cluster formation has 
been described for a wide variety of receptors, ranging from 
chemotactic receptors in bacteria to growth factor and neurotransmitter 
receptors in mammalian cells, a mechanistic understanding of 
the underlying molecular processes is still puzzling. In an attempt 
to fill this gap we followed a combined experimental and theoretical 
approach by dissecting and modulating cargo binding, internalization 
and cellular response mediated by KDEL receptors (KDELRs) at the 
mammalian cell surface after interaction with a model cargo/ligand. 
Using a fluorescent variant of ricin toxin A chain as KDELR-ligand 
(${\scriptstyle\text{eGFP-RTA}}^{\scriptscriptstyle\text{H/KDEL}}$), 
we demonstrate that cargo binding induces dose-dependent receptor 
cluster formation at and subsequent internalization from the membrane 
which is associated and counteracted by anterograde and 
microtubule-assisted receptor transport to preferred docking 
sites at the plasma membrane. By means of analytical arguments 
and extensive numerical simulations we show that cargo-synchronized 
receptor transport from and to the membrane is causative for 
KDELR/cargo cluster formation at the mammalian cell surface.  
\end{abstract}

\begin{document}

\flushbottom

\maketitle

\thispagestyle{empty}


\dropping[0pt]{2}{S}ensing of and responding to extracellular stimuli is an 
intrinsic property of eukaryotic cells to tightly regulate essential 
basic processes such as proliferation, migration, neurotransmission, 
or even immune defense \cite{Park15,Munoz08,Wajant03,Renner08,Jaqaman12,
Casaletto12}. In particular plasma membrane (PM) receptors, e.g.\ G-protein 
coupled receptors (GPCRs), play an important role in recognizing 
extracellular ligands, such as peptide hormones or drugs, and subsequently 
transducing the exogenous signal into a cellular response \cite{Hanlon15}. 
In this context, a series of cell surface receptors, including EGF and 
T-cell receptors as well as receptors that are parasitized by certain 
A/B toxins or viruses for endocytic internalization, are known 
to cluster in dynamic membrane nano-domains allowing cells to tune 
signaling efficiency and ligand sensitivity, or control protein 
interactions \cite{Hanlon15,Abulrob10,Bray98,Abrami03,Eierhoff10,
Gao05}. Since various human diseases are directly linked to 
abnormalities in membrane-receptor distribution and/or activation, 
it is important to understand the underlying mechanistic principles 
responsible for receptor clustering and dynamic reorganization to 
develop potential strategies for a therapeutic treatment 
\cite{Casaletto12,Abulrob10,ElRayes04}.

To address such essential biophysical aspects in receptor biology, 
we focused on mammalian KDEL receptors (KDELRs) at the cell surface 
that we and others have shown to be responsible for the sensing and 
binding of KDEL-cargo and KDEL-bearing A/B toxins \cite{Riffer02,
Eisfeld00,Schmitt06,Henderson13}. Besides having a central function 
in the retrieval of luminal proteins of the endoplasmic reticulum 
(ER) and in KDEL-cargo uptake from the cell surface, KDELRs are also 
known to act as GPCRs in the regulation of gene expression. The 
loss of KDELR1 has been recently demonstrated to cause lymphopenia 
and a failure in controlling chronic viral infections \cite{Siggs15,
Cancino14,Semenza90}. Because of the biomedical importance of 
KDELRs at the mammalian cell surface we addressed this aspect in 
more detail and aimed to answer the following questions: (i) How 
are KDELRs distributed in the PM and how does cargo binding affect 
receptor dynamics and distribution at the cell surface? (ii) How 
do cells respond to cargo binding and what is the underlying 
cellular mechanism? In contrast to the majority of studies on 
receptor clustering that either focused on biological or on 
theoretical aspects, we here followed a combined experimental, 
computational, and theoretical approach to dissect and modulate 
cargo binding, internalization and cellular response mediated by 
KDELRs at the mammalian cell surface. We thereby demonstrate that 
cargo binding induces dose- and temperature-dependent receptor 
clustering at and internalization from the PM that is accompanied 
and counteracted by microtubule-assisted anterograde receptor 
transport to distinct docking sites at the membrane. Based on 
the results of extensive Monte Carlo simulations and analytical 
arguments we disentangle the effects of surface dynamic processes 
from those of cargo-synchronized anterograde KDELR transport along 
the microtubule network towards and from the PM, and verify that the 
statistical properties and temporal evolution of the receptor 
cluster-size distribution is mainly induced and controlled by 
the later process.

\section*{Results}
\label{sec:Results}
KDELRs represent transmembrane proteins which recognize and 
bind soluble residents of the ER containing a C-terminal retention 
motif (KDEL or KDEL-like) to prevent escape from the secretory 
pathway \cite{Semenza90,Pelham88}. Recent studies however 
demonstrated that KDELRs are not restricted to ER and Golgi 
compartments but also localize in the PM where they bind 
KDEL-cargo such as mesencephalic astrocyte-derived neurotrophic 
factor (MANF) \cite{Henderson13} and internalize microbial A/B 
toxins such as the HDEL-bearing K28 virus toxin \cite{Riffer02,
Eisfeld00,Schmitt06}. Until now, however, it is unknown what 
mechanistically happens after a potential H/KDEL-cargo has bound 
to the pool of PM localized KDELRs. In addition to the equilibrium 
between anterograde receptor delivery to and internalization from 
the plasma membrane, receptor clustering as well as lateral membrane 
diffusion in response to ligand binding could play a key role 
in determining the total amount of KDELRs at the cell surface, 
similar to how EGFR (epidermal growth factor receptor) and AChR 
(acetylcholine receptor) control ligand sensitivity and activate 
signaling pathways \cite{Abulrob10,Bray98,Hezel10}. 

\begin{figure*}
\centerline{\includegraphics[width=0.85\textwidth]{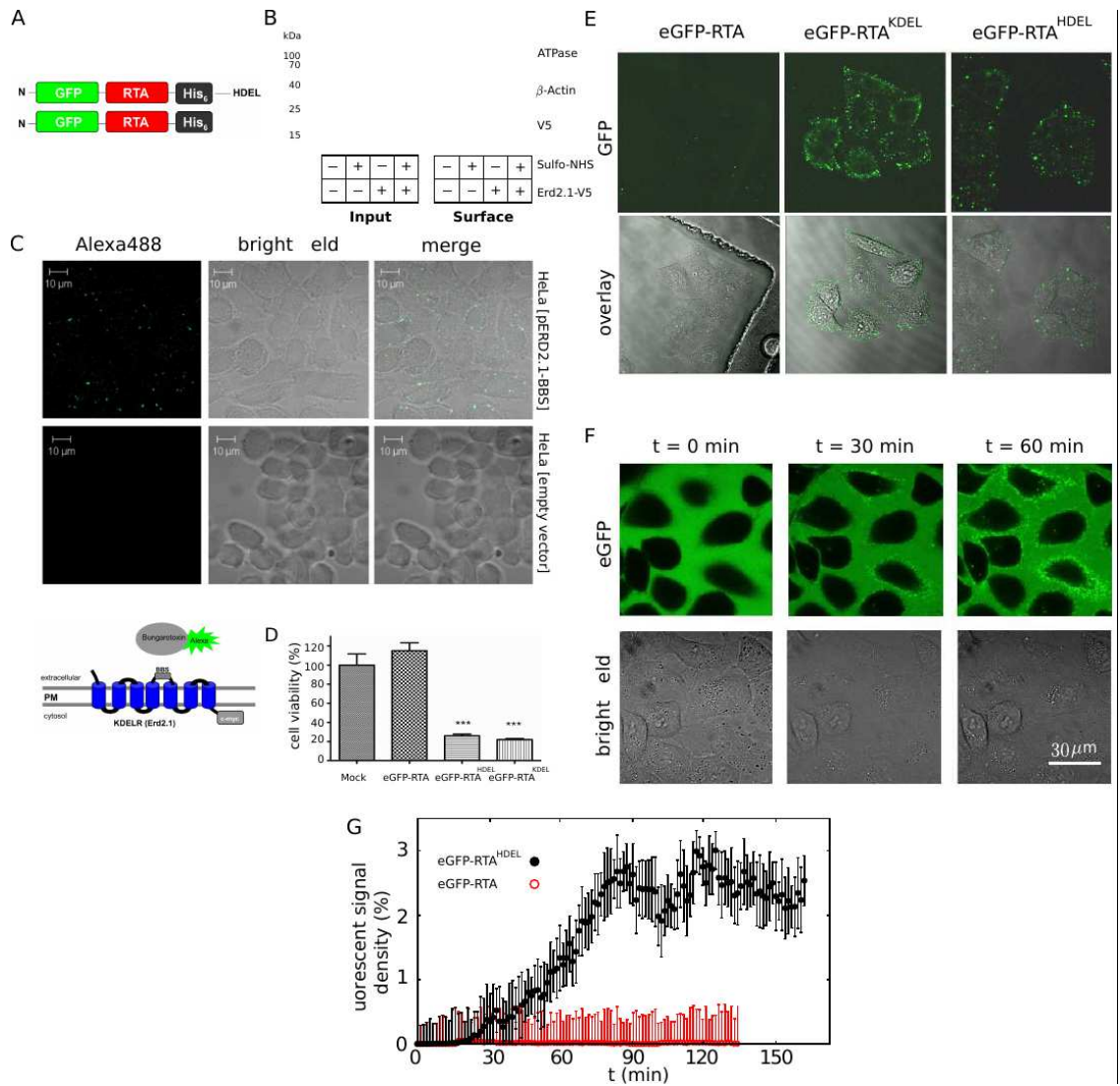}}
\caption{H/KDEL-cargo binding to the mammalian cell surface induces 
receptor cluster formation. (A) (top) Schematic outline of the 
fluorescent model cargo ${\scriptstyle\text{eGFP-RTA}}^{
\scriptscriptstyle\text{HDEL}}$ consisting of the cytotoxic 
A-subunit of ricin (RTA), mammalian enhanced GFP (eGFP) and a 
C-terminal $\text{(His)}_6$-Tag for purification. (bottom) eGFP-RTA 
lacking a KDELR binding motif served as negative control. (B) Cell 
surface biotinylation of mammalian KDELR1. HeLa cells were transiently 
transfected with KDELR1 (Erd2.1-V5 (+)) or an empty vector (-) 
and cultivated for $\text{48\,h}$. Cell surface proteins were 
biotinylated by treatment with (+) or without (-) Sulfo-NHS-SS-Biotin 
and purified with streptavidin beads. Whole cell lysates (input) 
served as control to determine the total amount of Erd2.1-V5 
(detected with anti-V5), while $\beta$-actin and Na+/K+ ATPase 
served as cytosolic and plasma membrane marker proteins, 
respectively. Membrane fraction (surface) illustrates the 
total fraction of proteins at the cell surface. (C) (bottom) 
Schematic outline of $\alpha$-bungarotoxin (Btx) cell surface binding. 
HeLa cells expressing a KDELR variant in which a Btx binding site 
(BBS) was inserted between positions T114 and P115 of c-myc-tagged 
KDELR1 (Erd2.1) were treated with Alexa488-labeled $\alpha$-Btx. As Btx 
is incapable to cross the mammalian PM, any physical interaction 
between Btx and BBS can only occur if KDELR1 is present in the PM. 
(top) Confocal laser scanning microscopy of HeLa cells transfected 
with pERD2.1-BBS-cmyc or an empty vector control and treated with 
10 $\mu\text{g/ml}$ Alexa488-labeled $\alpha$-Btx. (D) In vivo toxicity 
of ${\scriptstyle\text{eGFP-RTA}}^{\scriptscriptstyle\text{H/KDEL}}$ 
against HeLa cells. Cell viability ($\text{N{=}3}$, $\text{n{=}5}$ 
replicates) was determined after $\text{48\,h}$ in the presence or 
absence of 160 $\mu\text{g/ml}$ of the indicated RTA variant (Mock, 
PBS buffer). Mean values and standard deviations are displayed 
(${*}{*}{*}, P{<}0.001$, t test). (E) Fluorescence microscopy of cargo 
binding at the cell surface. HeLa cells were treated with 160 
$\mu\text{g/ml}$ ${\scriptstyle\text{eGFP-RTA}}^{\scriptscriptstyle\text{H/KDEL}}$ 
or ${\scriptstyle\text{eGFP-RTA}}$ for $\text{5 min}$ and cargo 
binding was analyzed after 10 washing steps. (F) Live cell imaging 
(45 frames/h) of HeLa cells treated with 160 $\mu\text{g/ml}$ 
${\scriptstyle\text{eGFP-RTA}}^{\scriptscriptstyle\text{HDEL}}$. 
Three representative time points ($\text{0, 30, 60 min}$) are shown. 
(G) Temporal evolution of the density of cargo signals at the 
surface of HeLa cells. The accumulation of fluorescent signals 
is shown after treatment with 160 $\mu\text{g/ml}$ ${\scriptstyle
\text{eGFP-RTA}}^{\scriptscriptstyle\text{HDEL}}$ or ${\scriptstyle
\text{eGFP-RTA}}$. The symbols represent the optimal signal-to-noise 
ratio in image analysis. The error bars reflect the variation 
range of signal intensity for different threshold values of 
image analysis parameters. The functional form only weakly depends 
on the choice of the threshold values.}
\label{Fig:1}
\end{figure*}

\subsection*{Design and biological activity of a model KDELR cargo}
KDELR cluster formation at the mammalian cell surface in response to 
cargo binding was analyzed and visualized on a model cargo by 
using a GFP-tagged variant of the cytotoxic A-subunit of ricin (RTA) 
extended by a C-terminal H/KDEL motif (Fig.\,\ref{Fig:1}A). Using 
cell surface biotinylation, we were able to detect KDELR1 at 
the cell surface of mammalian cells (Fig.\,\ref{Fig:1}B), in 
agreement with recent studies in which a pool of KDELR1 was observed 
at the PM of neuroblastoma cells \cite{Henderson13}. Cell surface 
localization of KDELR1 was also analyzed by adapting an imaging 
assay originally designed to confirm PM-localization of AMPA 
receptors through binding of the snake venom $\alpha$-bungarotoxin, 
Btx \cite{Balass97,Sekine-Aizawa04}. For imaging analysis at the 
cell surface, a Btx binding site (BBS) was inserted into an 
extracellular loop of mammalian Erd2.1 (KDELR1) and subsequently 
used to visualize physical Btx/KDELR interaction at the PM 
(Fig.\,\ref{Fig:1}C, bottom). Extracellular binding of Alexa488-labeled 
Btx to the modified and in vivo functional KDELR1 variant containing 
an Btx binding motif in an extracellular loop of the receptor 
further supported the biotinylation data and likewise demonstrated 
that a minor but significant number of receptors localizes in 
clusters at the PM of HeLa cells (Fig.\,\ref{Fig:1}C, top).
The recombinantly expressed and purified cargo protein 
${\scriptstyle\text{eGFP-RTA}}^{\scriptscriptstyle\text{H/KDEL}}$ 
remained toxic to HeLa cells (see Fig.\,\ref{Fig:1}D and 
Supplementary Fig.\,S1), confirming earlier observations 
that the addition of a C-terminal H/KDEL motif to RTA enhances its 
in vivo toxicity \cite{Wales93,Becker11}. In contrast to the negative 
control of eGFP-RTA lacking a C-terminal KDELR binding site, 
${\scriptstyle\text{eGFP-RTA}}^{\scriptscriptstyle\text{H/KDEL}}$ 
rapidly bound to the cell surface within seconds to minutes, 
indicating that a fraction of KDELRs at the PM is responsible 
for ligand binding (Figure \ref{Fig:1}E and Supplementary 
Fig.\,S2). This is further supported by the similar 
behavior seen in cell surface cargo clustering in response to 
${\scriptstyle\text{eGFP-RTA}}^{\scriptscriptstyle\text{HDEL}}$ 
addition and KDELR1 pattern formation after Btx binding 
(Fig.\,\ref{Fig:1}C). As analogous KDELR1/cargo clustering 
was also observed in different cell lines (such as HEK-293T 
and RAW-Blue), KDELR-mediated cargo binding at the PM is not 
restricted to just a single cell type but rather seems a general 
phenomenon in mammalian cells (Supplementary Fig.\,S3). 
Immunostaining of non-permeabilized HeLa cells as well as 
binding studies at $4^{\circ}$C further demonstrated that 
${\scriptstyle\text{eGFP-RTA}}^{\scriptscriptstyle\text{HDEL}}$ 
signals are indeed present at the plasma membrane and, thus, not 
restricted to signals of internalized KDELR/cargo complexes 
(Supplementary Fig.\,S2B and Fig.\,S4). 
In addition, increased toxicity of ${\scriptstyle\text{eGFP-RTA}}^{
\scriptscriptstyle\text{H/KDEL}}$, visible intracellular fluorescent 
signals, especially after longer incubation ($>6$ h) in live cell 
imaging, and the observed endocytic uptake of KDELR1 in the 
modified/reversed biotinylation experiment further indicate that 
${\scriptstyle\text{eGFP-RTA}}^{\scriptscriptstyle\text{H/KDEL}}$ 
is indeed internalized from the mammalian cell surface(see 
Fig.\,\ref{Fig:1}D, \ref{Fig:1}G, Supplementary Fig.\,S5 and 
Supplementary Movies S1 and S2).   

Furthermore, live cell imaging of cells loaded with ${\scriptstyle
\text{eGFP-RTA}}^{\scriptscriptstyle\text{HDEL}}$ (see Fig.\,\ref{Fig:1}F) 
identified a strict time-dependent accumulation of fluorescent cargo signals 
at the PM which was absent in control cells treated with eGFP-RTA 
lacking a KDELR binding motif (Fig.\,\ref{Fig:1}H). Interestingly, 
the development of fluorescent signals/clusters of ${\scriptstyle
\text{eGFP-RTA}}^{\scriptscriptstyle\text{HDEL}}$ at the cell surface 
occurred in distinct phases: Initially, the system remained relatively 
inactive for a short time ($t{<}\text{20 min}$). After this transient 
regime, an exponential growth was observed, which eventually saturated 
at $t{>}\text{80 min}$. The observed huge fluctuations of the accumulated 
receptor density at the PM is a signature of the stochasticity of 
the underlying nonequilibrium process, where the system ultimately 
reaches a balance between the loss of surface receptors due to 
endocytosis and gain by recycling them.

\begin{figure*}
\centerline{\includegraphics[width=0.9\textwidth]{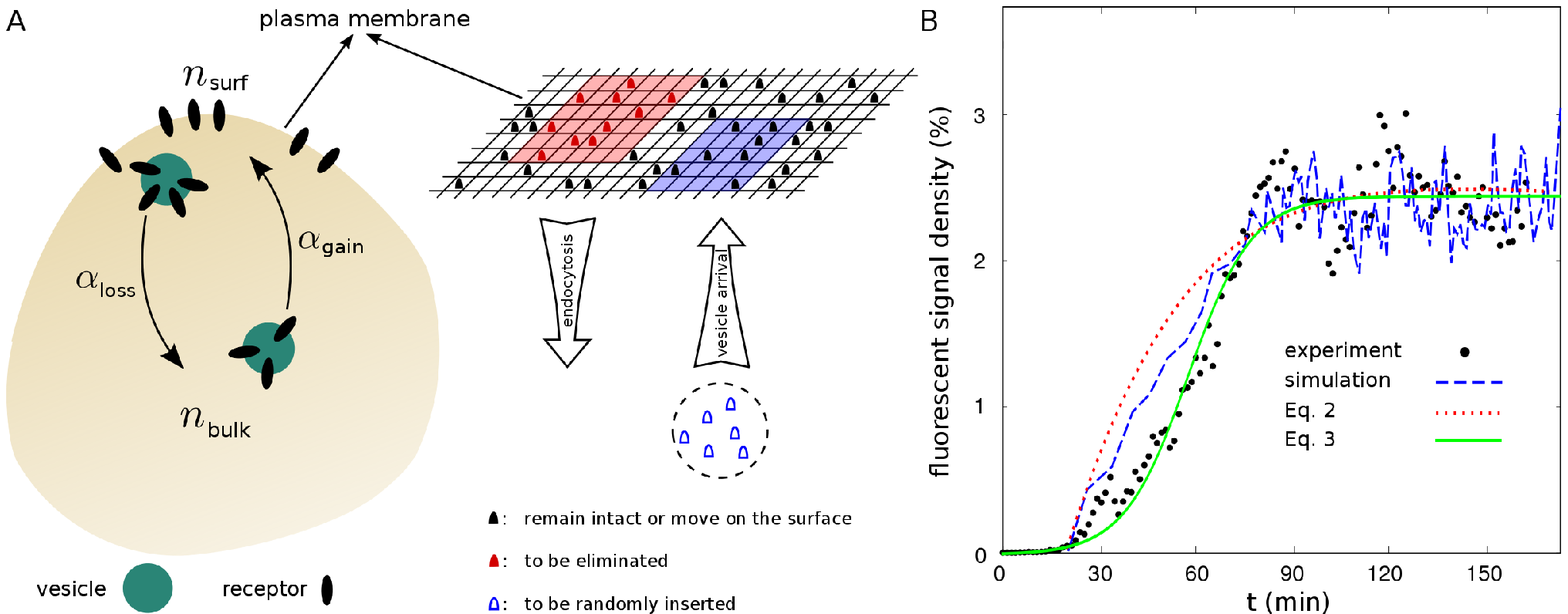}}
\caption{(A) Schematic representation of (left) the minimal model of 
receptor cycle between the PM and endosomes, and (right) the simulation 
method. An example of a randomly chosen area for endocytosis (vesicle 
arrival) is marked in red (blue). Possible scenarios for the evolution 
of the surface receptor population during the next simulation step 
are shown. (B) Time evolution of the density of accumulated cargo at 
the cell surface. A comparison is made between experimental data, 
simulation results (a single realization), and the analytical 
expressions Eqs.\,\ref{Eq:Density} and \ref{Eq:Density2}. The dotted 
line indicates the analytical prediction via Eq.\,\ref{Eq:Density} 
for $\alpha_{_\text{gain}}{=}\alpha_{_\text{loss}}{=}1.3{\times}
10^{-4}\text{s}^{-1}$. The starting time of simulations and analytical 
expression 2 is shifted to take into account the initial inactive 
regime in experiments.}
\label{Fig:2}
\end{figure*}

\subsection*{Adaptive Monte Carlo simulations of KDELR/cargo dynamics 
at the cell surface}
Aiming at better understanding the in vivo observed KDEL/cargo 
interactions at the cell surface, we performed extensive Monte Carlo 
simulations which shed light on the underlying mechanisms of 
receptor clustering at the cell membrane. We modeled the cell 
membrane as a lattice with a spacing of the size of receptors 
(${\sim} \text{10 nm}$) \cite{Reddy10,Guo99} and periodic boundary 
conditions (Fig.\,\ref{Fig:2}A). Each lattice site can be occupied 
by at most one receptor which is either liganded or unliganded. The 
membrane size ($4{\times}10^6$ lattice sites) is comparable to that 
of a typical HeLa cell. Assuming that the frequency, spatial extent, 
and target region of endocytosis and recycling of receptors are 
independent stochastic events, we introduced asymmetric rates of 
endocytosis and vesicle arrival, and chose a random target region 
on the membrane for each event. Considering the normal size of 
clathrin-coated vesicles to be in the range of 50 to 100 nm 
\cite{Agrawal10}, we allowed the extent of events to vary within 
$5{\times}5$ to $10{\times}10$ lattice sites. An endocytosis 
event leads to elimination of all receptors within the affected 
zone. The number of receptors carried by an incoming vesicle 
was chosen randomly from $0$ to the maximum capacity of that 
vesicle and distributed randomly within the targeted zone on 
the PM upon availability of empty sites. One may also switch 
on the receptor surface dynamics, including lateral membrane 
diffusion and receptor-receptor interactions. Starting with 
an initial random configuration of receptors on the lattice, 
the surface density evolves and finally reaches a nonequilibrium 
steady state by balancing the receptor gain and loss. There 
are density fluctuations in the steady state due to the 
stochastic origin of endocytosis and vesicle arrival events. 
As shown in Fig.\,\ref{Fig:2}B, the experimental data could be 
qualitatively reproduced in simulations by tunning the initial 
density at the PM and the gain and loss rates. Notably, the 
amplitude of steady-state oscillations in simulations obtained 
for a single realization is comparable to the experimental data.

\subsection*{Models for receptor cycle}
We first developed a minimal theoretical model for loss and gain 
of receptors during endocytosis and recycling back to the surface.
Assuming that the total number of receptors in the cell is 
conserved within our experimental time window, the fractions of 
receptors at the cell surface $n_{_\text{surf}}$ and inside the 
cell $n_{_\text{bulk}}$ are related as $n_{_\text{surf}} {+} 
n_{_\text{bulk}}{=}1$. Denoting the endocytosis and vesicle 
arrival rates, respectively, with $\alpha_{_\text{loss}}$ and 
$\alpha_{_\text{gain}}$, the evolution of the average fraction 
of receptors at the plasma membrane $n_{_\text{surf}}(t)$ in a 
simple form can be described as 
\begin{equation}
\frac{\text{d}n_{_\text{surf}}}{\text{dt}}=\alpha_{_\text{gain}} \, 
n_{_\text{bulk}} -\alpha_{_\text{loss}} \, n_{_\text{surf}}.
\label{Eq:DensityMasterEq}
\end{equation}
Denoting the initial and steady-state fractions of surface receptors 
with $n^{\text{o}}_{_\text{surf}}$ and $n^{\infty}_{_\text{surf}}$, 
one obtains $n^{\infty}_{_\text{surf}}{=}\frac{\alpha_{_\text{gain}}}{
\alpha_{_\text{loss}}{+}\alpha_{_\text{gain}}}$, and the time evolution 
of the average fraction of surface receptors follows 
\begin{equation}
\displaystyle n_{_\text{surf}}(t) = n^{\infty}_{_\text{surf}} 
+ (n^{\text{o}}_{_\text{surf}}{-}n^{\infty}_{_\text{surf}}) \, 
\text{e}^{-t/\tau\!_{_\text{o}}},
\label{Eq:Density}
\end{equation}
with the characteristic time $\tau\!_{_\text{o}}{=}\frac{1}{
\alpha_{_\text{gain}}{+}\alpha_{_\text{loss}}}$. Thus, $n^{
\infty}_{_\text{surf}}$ and $\tau\!_{_\text{o}}$ are controlled 
by the rates $\alpha_{_\text{loss}}$ and $\alpha_{_\text{gain}}$. 
Indeed, the cycle of receptors in the cell is more complicated; 
there are more influential parameters involved and the receptor 
conservation assumption does not hold in general. Nevertheless, 
as shown in Fig.\,\ref{Fig:2}B, Eq.\,\ref{Eq:Density} 
qualitatively reproduces the trends observed in experiments, 
though the curvature change is not captured. 

The previous approach would reflect a situation where the 
transport of receptors to and from the plasma membrane is 
neither influenced by exclusion nor by self-amplification. 
However, if one assumes that binary excluded-volume interactions 
between receptors  have to be considered and/or self-amplification 
of the receptor transport plays a role, the evolution of the 
fraction of surface receptors can be roughly described as 
\begin{equation}
\frac{\text{d}n_{_\text{surf}}}{\text{dt}} =  A + B n_{_\text{surf}} 
- C n^2_{_\text{surf}}.
\label{Eq:Density2}
\end{equation}
By fitting the free parameters, the above equation captures 
quantitatively well the in vivo observed dynamics over the 
whole time window (solid line in Fig.\,\ref{Fig:2}B).

\subsection*{Effect of temperature and ligand concentration on KDELR 
cluster formation}
Prior to experimentally investigating if temperature changes affect 
KDELR/cargo clustering at the cell surface and follow the van-`t-Hoff'sche 
rules, we assumed that KDELR clustering at $25^{\circ}$C should slow 
down by a factor of 2-4 without changing the overall kinetics and 
shape of the curve. Based on this assumption, we reduced the endocytosis 
and vesicle arrival rates in simulations by a factor of $3$. The results 
predicted a timely retardation of KDELR membrane cluster formation at 
$25^{\circ}$C while the overall saturation level was similar to the 
one reached at $37^{\circ}$C (Fig.\,3A, left panel). Since the 
experimental data nicely confirmed the numerical predictions 
(Fig.\,3A, right panel), it can be concluded that KDELR/cargo cluster 
formation is a temperature-dependent process. It can be also deduced 
from the minimal analytical model presented in the previous section, 
that the final saturation level $n^{\infty}_{_\text{surf}}$ remains 
invariant under a symmetric scaling of the rates (i.e.\ $\alpha_{_
\text{loss}}{\rightarrow}\,\kappa\alpha_{_\text{loss}}$ and 
$\alpha_{_\text{gain}}{\rightarrow}\,\kappa\alpha_{_\text{gain}}$), 
while the characteristic relaxation time is rescaled as $\tau\!_{_
\text{o}}{\rightarrow}\,\tau\!_{_\text{o}}/\kappa$.

To determine any effect of cargo concentration on cluster formation 
at the PM, HeLa cells were treated with different doses of ${\scriptstyle
\text{eGFP-RTA}}^{\scriptscriptstyle\text{HDEL}}$; the corresponding 
results revealed a strict dose-dependency of KDELR/cargo cluster 
formation at the cell surface (Fig.\,3B). In contrast to the impact 
of temperature, the variation of the saturation level with changing 
the cargo concentration indicated that the endocytosis and vesicle 
arrival rates are differently affected, i.e.\ they scale as 
$\alpha_{_\text{loss}}{\rightarrow}\,\kappa\,\alpha_{_\text{loss}}$ 
and $\alpha_{_\text{gain}}{\rightarrow}\,\kappa'\alpha_{_\text{gain}}$ 
with $\kappa{\neq}\kappa'$. This is indeed necessary to reproduce 
the experimental data in simulations (Fig.\,3B, left panel). Denoting 
the steady-state fraction of surface receptors at low and high 
concentrations with $n^{\infty}_{_\text{low}}$ and $n^{\infty}_{_
\text{high}}$ ($n^{\infty}_{_\text{high}}{>}n^{\infty}_{_\text{low}}$), 
from the minimal theoretical model one finds that $\displaystyle
\frac{\alpha_{_\text{gain}}^{\text{high}}}{\alpha_{_\text{gain}}^{
\text{low}}}{>}\frac{\alpha_{_\text{loss}}^{\text{high}}}{\alpha_{_
\text{loss}}^{\text{low}}}$, thus, the vesicle arrival rate is more 
sensitive to the concentration changes than the endocytosis rate. 
Interestingly, reduction of the saturation level at lower cargo 
concentrations suggests that mammalian cells can somehow modulate 
the response depending on extracellular ligand concentration. They 
are capable to sense the actual concentration of cargo binding and 
subsequently regulate the total amount of KDELRs at the cell surface. 
Hence, the combination of experimental data and numerical results 
provides a first mechanistic insight into KDELR/cargo clustering 
at the mammalian cell surface.  

\begin{figure}[t]
\centerline{\includegraphics[width=0.48\textwidth]{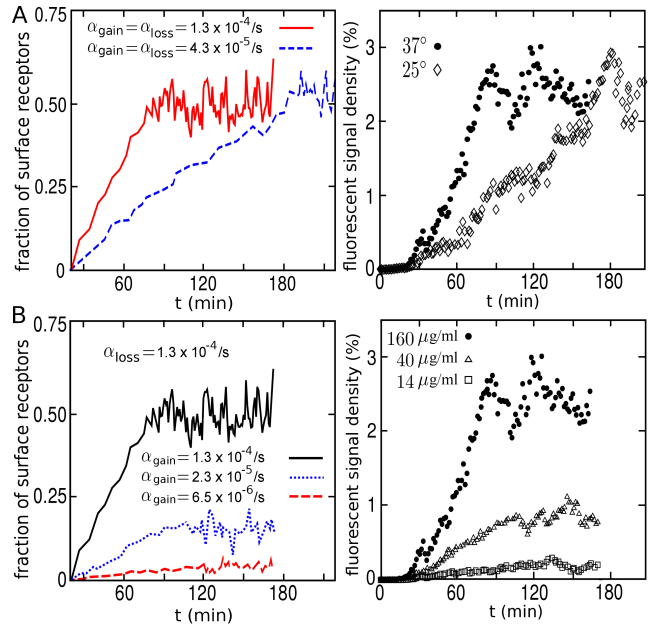}}
\caption{KDELR/cargo clustering is dose- and temperature-dependent. 
(A) Changes in cargo accumulation of ${\scriptstyle\text{eGFP-RTA}}^{
\scriptscriptstyle\text{HDEL}}$ at the surface of HeLa cells cultivated 
at $25^{\circ}$C or $37^{\circ}$C. The 3-fold reduced activity in 
simulations (left) represents the known effect of temperature on 
intracellular transport processes (e.g.\ endocytosis and exocytosis). 
(right) The experimental results at $25^{\circ}$C and $37^{\circ}$C. 
(B) Effect of changing the concentration of the model cargo 
${\scriptstyle\text{eGFP-RTA}}^{\scriptscriptstyle\text{HDEL}}$ 
on KDELR/cargo clustering at the plasma membrane. The indicated 
rates in simulations (left) were adopted to obtain the best fits 
to the experimental data (right).}
\label{Fig:3}
\end{figure}

\subsection*{Anterograde KDELR transport to preferential plasma membrane 
arrival sites}
The cellular plasma membrane resembles a thoroughly regulated and highly 
dynamic compartment that contains cell surface micro-domains like lipid 
rafts or caveolea \cite{Hancock06}. It is well documented that plasma 
membrane receptors such as AchR, EGFR or TGF-$\beta$ are associated with 
lipid rafts \cite{Abulrob10,Chen15,Campagna06} and that preferential receptor 
cluster formation in distinct micro-domains of the PM provides an 
important means to regulate downstream signaling as shown for EGFR 
\cite{Abulrob10}. Moreover, it has been proposed that T cell receptor 
pre-clustering at the cell surface contributes to a significant increase 
in ligand sensitivity and accelerates signaling pathway activation 
\cite{Castro14}.  

To understand how KDELR/cargo clustering evolves and whether or not 
KDELRs are likewise arranged in receptor pre-clusters or micro-domains, 
cluster size distribution $P(s)$ of the model cargo ${\scriptstyle
\text{eGFP-RTA}}^{\scriptscriptstyle\text{HDEL}}$ was determined at 
different time points during the clustering process. As shown in 
Figs.\,4A and 4B, relatively small clusters were visible at early time points, 
while larger clusters appeared at longer times. A detailed analysis 
indicated that the growth of the largest cluster eventually stopped after 
reaching the stationary state. Notably, the cluster size distribution 
nearly followed a power-law decay $P(s){\sim}s^{-\beta}$ with a rather 
time-invariant exponent $\beta{\approx}2$ throughout the clustering 
process, except for the initial transient regime ($t{<}\text{20 min}$). 

The functional form of the cluster size distribution $P(s)$ indeed 
implies which of the underlying mechanisms of receptor dynamics is 
dominant. In Monte Carlo simulations, we first assumed that the target 
zones for endocytosis and vesicle arrival events are randomly chosen, 
without allowing the distributed receptors to move on the surface.  
Next, we examined the main possible scenarios for receptor surface 
dynamics including lateral diffusion on the membrane \cite{Boggara13,
Saffman75} and receptor-receptor attraction \cite{Reddy10,Guo99} 
(both with the short range of nearest-neighbor sites). Figure 
4C shows that none of the resulting aggregation patterns 
was capable of producing a power-law size distribution; the tails 
of the resulting distributions rather follow an exponential-like 
decay. The algebraic form can be recovered under the assumptions 
that KDELRs have distinct and preferred arrival sites at the plasma 
membrane and the transport is self-amplified. This was achieved 
in the following way. The spatial distribution of targeting 
probability was changed from a uniform to a multiple-peaked 
Gaussian one. The peaks represent the places where MTs approach 
the cell cortex and distribute their vesicles, which are supposed 
to diffuse on the actin filament network until they finally reach 
the membrane. Additionally, the surface was divided into subdomains 
obtained by Voronoi tessellation of area around each peak, and 
a newly generated vesicle choses a target subdomain with a 
probability proportional to the transport-activity history 
of the corresponding MT. 

Let us consider a rather simple process of receptor aggregation, 
in which the time evolution of the probability $P(s)$ of having 
receptor clusters of size $s$ is expressed via the master equation 
$\frac{dP(s)}{dt} = \sum_{i{+}j{=}s}\frac12 P(i) P(j){-} \sum_{i} 
P(i) P(s)$. The gain and loss terms on the right-hand side account 
for creation of $s$-size clusters from the coalescence of two smaller 
ones of sizes $i$ and $j$, and merging of $s$-size clusters with other 
ones, respectively. Starting from an initial configuration with 
randomly distributed single receptors, the master equation can be 
recursively solved to get $P(s){\sim}\exp[-\ln(2{+}\frac2t)s]$, 
i.e.\ $P(s)$ decays exponentially with a time-dependent exponent. 
At long times, the slope decreases and $P(s)$ evolves towards a 
flat distribution. Our attempts to consider more complicated 
aggregation scenarios, such as introducing diffusion or aggregation 
with input, failed to simultaneously reproduce the power-law 
and time-invariant features of $P(s)$. In contrast, it can be 
verified that preferential attachment of receptors to the existing 
clusters reproduces the experimentally observed distribution. We 
consider a simple clustering process in which a new receptor 
is added to the surface at each time step, and it attaches to 
the cluster of size $s_i$ with a probability $p_{s_i}$ which is 
proportional to the cluster size, i.e.\ $\displaystyle p_{s_i}
{=}\frac{s_i}{\sum_j s_j}$. The sum runs over all clusters, thus, 
reflects the total number of receptors and grows linearly with 
time. The rate at which $s_i$ changes can be assumed to be 
proportional to $p_{s_i}$
\begin{equation}
\frac{\partial s_i}{\partial t}=p_{s_i}=\frac{s_i}{\sum_j s_j}=
\frac{s_i}{t}.
\label{Eq:si-rate}
\end{equation}
Denoting the initiation time of cluster $i$ with $t_i$, one obtains
$s_i(t)=t{/}t_i$. The probability that a cluster is smaller than 
$s$ is given as
\begin{equation}
P[s_i(t){<}s]=P(t_i{>}\frac{t}{s}).
\label{Eq:Pcumulative}
\end{equation}
The probability $P(t_i)$ of initiation at time $t_i$ has a 
constant probability density with respect to time, i.e.\ 
$P(t_i)=\frac{1}{t}$. Substituting this into 
Eq.\,\ref{Eq:Pcumulative} we get
\begin{equation}
P(t_i{>}\frac{t}{s})=1-\frac{t}{s}\cdot\frac{1}{t}=1-\frac{1}{s}.
\label{Eq:Pcumulative2}
\end{equation}
Finally, the cluster-size distribution can be obtained using
\begin{equation}
P(s)=\frac{\partial P[s_i(t){<}s]}{\partial s}=\frac{\partial}
{\partial s}(1-\frac{1}{s})=\frac{1}{s^2},
\label{Eq:Ps}
\end{equation}
which shows a power-law decay with a time-independent exponent $2$, 
in a remarkable agreement with the experimental results shown in 
Fig.\,4A. Since the receptor trafficking mainly occurs along 
MTs, vesicle exchange near the PM happens mostly in the vicinity 
of the regions where MTs approach actin filaments near the cell 
cortex. Our data are consistent with a feedback mechanism that 
amplifies receptor transport towards the plasma membrane in the 
presence of receptor clusters. 

\begin{figure*}
\centerline{\includegraphics[width=0.9\textwidth]{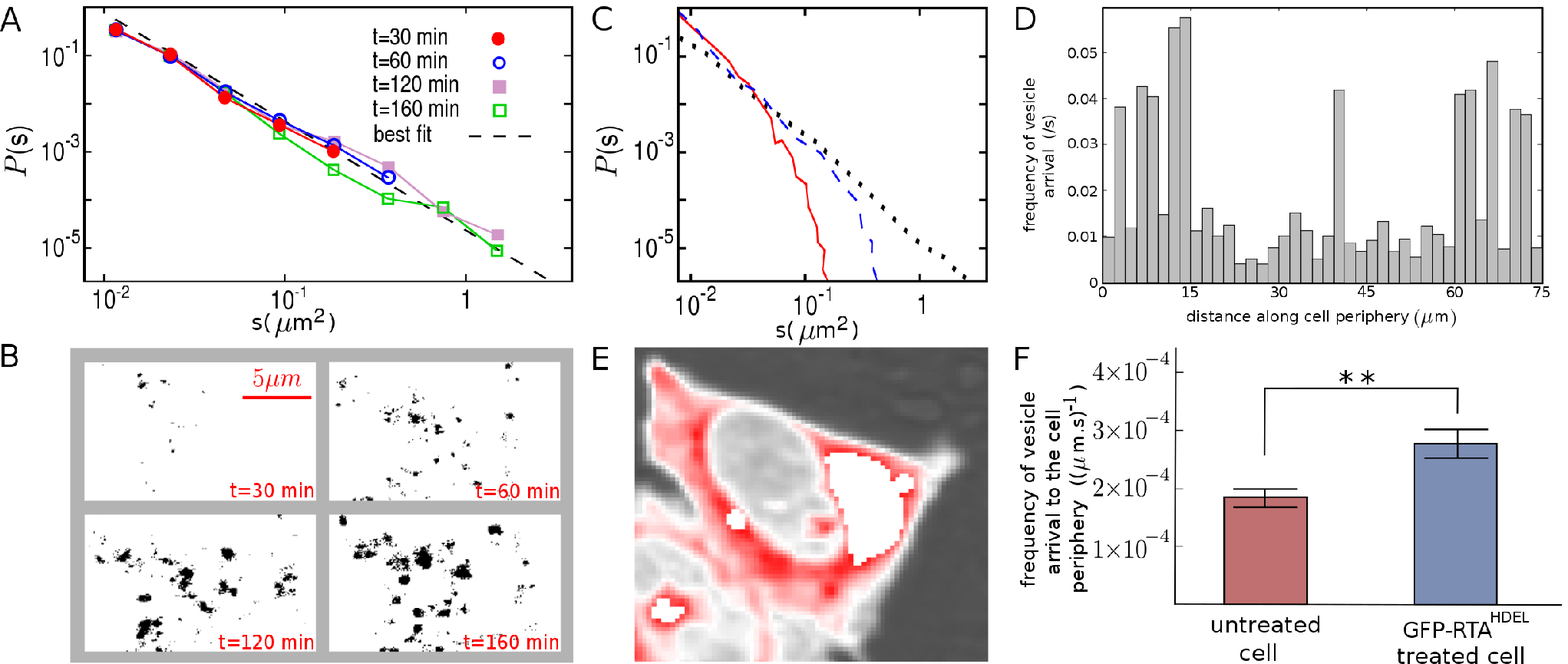}}
\caption{Preferential arrival sites of KDELRs at the plasma membrane. 
(A) The log-log plots of cluster-size distribution $P(s)$ of ${
\scriptstyle\text{eGFP-RTA}}^{\scriptscriptstyle\text{HDEL}}$ (160 
$\mu\text{g/ml}$) treated HeLa cells at the indicated time points. 
The dashed line corresponds to the best power-law fit $P(s){\sim}s^{-
\beta}$ with $\beta{\simeq}2.1{\pm}0.1$. (B) Evolution of the 
receptor clusters at the plasma membrane. A randomly chosen region 
of the cell surface is shown at different time points (see 
\emph{Suppl.~Info.} for the detailed description of the methodology of 
distinguishing the clusters and obtaining the cluster-size 
distribution). (C) A comparison of the resulting $P(s)$ from 
different receptor dynamic scenarios in simulations. The solid, 
dashed, and dotted lines denote the shape of $P(s)$ at 
$t{=}\text{120 min}$ for randomly distributed immobile 
receptors, aggregation process including lateral diffusion 
of receptors and nearest-neighbor attraction between them, and 
preferential attachment process, respectively. (D) The frequency 
of vesicle arrival at a sample cell periphery over a time 
window of 500 s. (E) In vivo dynamics of mCherry-ERD2.1. 
The transfected HeLa cells with mCherry-ERD2.1 were analyzed by 
CLSM (720 frames/h). The illustrated heat map represents the 
accumulated fluorescent signals of successive frames. The 
regions with high traffic load, e.g.\ around Golgi, are eliminated 
to provide a more clear color distinction near the cell surface. 
(F) The frequency of vesicle transport near the plasma membrane 
of untreated or ${\scriptstyle\text{eGFP-RTA}}^{\scriptscriptstyle
\text{HDEL}}$ treated cells. The data is averaged over bins of 
size $10\,\mu\text{m}$ (${*}{*}, P{\leq}0.01$, t test).}
\label{Fig:4}
\end{figure*}

Although the in vivo observed receptor dynamics is indeed more 
complicated and also depends on other factors (e.g. membrane 
thickness, lipid composition etc.) and involves both surface 
dynamic processes and the membrane-cytoplasm receptor cycle, 
our numerical and analytical findings suggest that the 
intracellular transport of vesicles along the microtubule 
network, which induces preferential zones for vesicle 
exchange at the PM, crucially controls the clustering at the 
mammalian cell surface. To experimentally prove this 
hypothesis, HeLa cells were transfected with mCherry-labeled 
KDELR1 (Erd2.1) and receptor dynamics was analyzed by live cell 
imaging (Supplementary Movie S3). By monitoring the frequency of 
vesicle arrival at the plasma membrane over a time window of 
500 s, it is shown in Fig.\,4D that anterograde KDELR transport 
is non-uniformly distributed along the plasma membrane, i.e.\ 
there are hot spots on the cell periphery which are targeted 
more frequently by the arrival of vesicles. The heat map of 
intracellular KDELR transport in Fig.\,4E illustrates the 
spatio-temporal distribution of KDEL receptors. Notably, 
a comparison between untreated and ${\scriptstyle\text{eGFP-RTA}}^{
\scriptscriptstyle\text{HDEL}}$ treated cells in Fig.\,4F 
indicates an increase in the transport rate in treated cells. 
Thus, the experimental findings support the numerical predictions 
and underline the importance of preferential absorption in 
regulating KDELR cluster formation. 

\subsection*{Microtubule and actin assisted receptor transport 
and membrane clustering}
Active protein transport along the cytoskeleton is mediated by 
actin filaments and microtubules (MTs). While filamentous F-actin 
is mainly localized in the cell cortex and involved in cell 
migration, endocytosis and vesicle-mediated cargo transport, 
MTs are responsible for dynein/kinesin driven active transport 
of vesicles and organelles \cite{Huber15,Hancock14,Soldati06}.

To verify that the MT-network is involved in intracellular 
KDELR trafficking, cells were co-transfected with GFP-tagged 
$\beta$-tubulin and mCherry-ERD2.1 and KDELR transport along 
MTs was visualized by CLSM. Despite the limited resolution of 
live cell imaging, directed transport of KDELR signals along 
MTs could be clearly observed (see e.g.\ Fig.\,5A), indicating 
that anterograde receptor transport is indeed MT based. A more 
quantitative analysis also revealed a relatively high probability 
of tubulin/KDELR signal co-localization (Fig.\,5B). Moreover, 
we observed that colchicine-mediated inhibition of MT assembly 
highly affects KDELR dynamics in vivo (Supplementary Movie S4A and S4B) 
and considerably reduces receptor clustering at the cell 
surface, see Fig.\,5C. We conclude that active receptor 
transport along MTs is a prerequisite for KDELR/cargo cluster 
formation at the plasma membrane. It has been shown \cite{Boggara13,
Keijzer11,Kharchenko07,Sakai91} that disruption of MTs causes 
the majority of EGFR or cAMPR clusters to be immobile, or affects 
the endocytosis of EGFRs; thus, MTs play a crucial role in 
organizing receptor clusters at the plasma membrane. 
Interestingly, KDELR mobility at the cell periphery was 
not completely blocked in colchicine treated cells indicating 
that cortical actin filaments which are unaffected by the drug 
are involved in and responsible for the observed KDELR endocytosis 
from the PM. Consistently, phalloidin-mediated actin inhibition 
caused a severe impairment of KDELR cluster development at 
the cell surface (Fig.\,5C, inset).  

\begin{figure*}
\centerline{\includegraphics[width=0.9\textwidth]{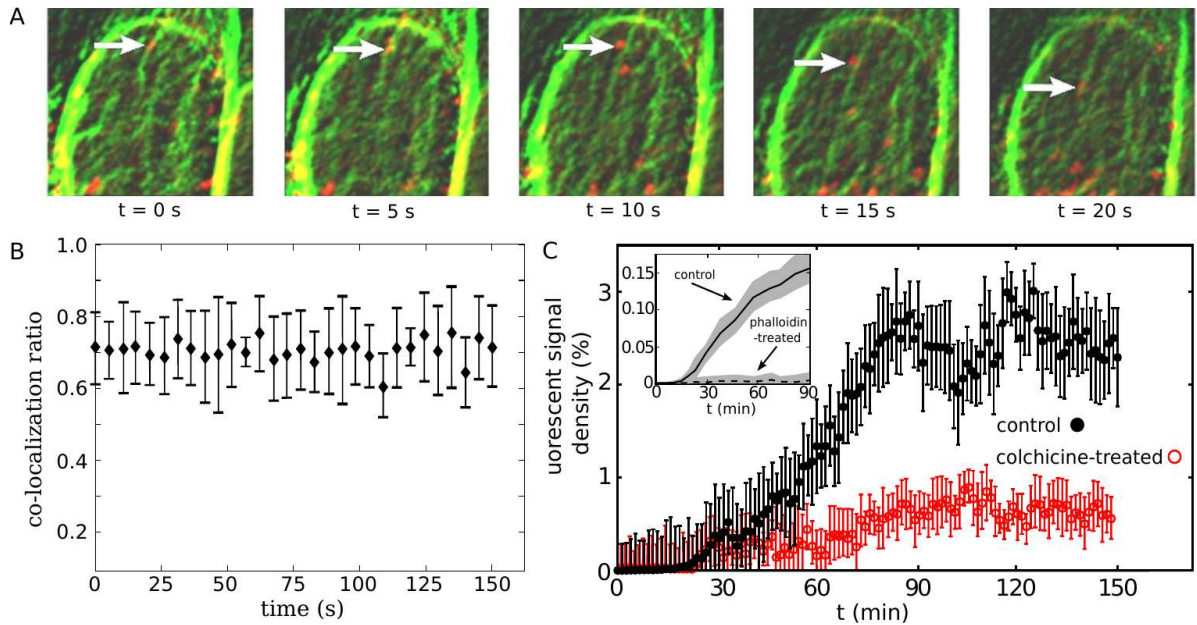}}
\caption{Microtubule-assisted KDELR transport is required for cargo 
clustering at the cell surface. (A) Tracking of single KDELR clusters 
(red) moving along the microtubule network (green). A sequence of 
five successive live cell imaging pictures (720 frames/h) of HeLa 
cells expressing mCherry-tagged Erd2.1 and GFP-tagged $\beta$-tubulin 
is shown. The arrows indicate an example of tubulin/KDELR signal 
co-localization. (B) Co-localization of GFP-tubulin and mCherry-Erd2.1. 
The ratio of correlated tubulin and KDELR pixels of the live cell 
imaging experiment is shown during 150 s. (C) KDELR/cargo cluster 
formation in colchicine (red) and phalloidin (inset) pre-treated 
cells ($2.5 \,\mu\text{M}$ colchicine, 60 min or $10 \,\mu\text{M}$ 
phalloidin, 90 min) after incubation with 160 $\mu\text{g/ml}$ 
${\scriptstyle\text{eGFP-RTA}}^{\scriptscriptstyle\text{HDEL}}$. 
Temporal evolution of the accumulated KDELR/cargo is compared 
with the untreated control cells. The inset shows a comparison 
between untreated (solid line) and phalloidin-treated HeLa cells 
(dashed line).}
\label{Fig:5}
\end{figure*}

\section*{Discussion}
\label{sec:Discussion}
Until recently, cargo recognition by KDEL-receptors has been 
assumed to mainly occur within the Golgi complex during retrograde 
transport of soluble ER residents back to the ER \cite{Pelham88,
Majoul01}. More recent studies, however, indicated that KDELRs 
are also responsible for cell surface binding of extracellular 
ligands such as the neurotrophic factor MANF or the microbial 
A/B toxin K28 \cite{Riffer02,Eisfeld00,Schmitt06,Henderson13}. 
Based on these findings we postulated that a model cargo containing 
a C-terminal H/KDEL amino acid motif and receptor binding site 
should likewise bind to cells and, thus, be suitable to track 
and analyze cargo binding and subsequent cellular responses. 
Using this approach we now demonstrate that treatment of cells 
with a fluorescent variant of RTA extended by a C-terminal H/KDEL 
motif is required and sufficient to promote specific cargo binding 
and clustering at the mammalian cell surface. Based on the 
experimental data presented here it can be deduced that the 
initial binding of ${\scriptstyle\text{eGFP-RTA}}^{\scriptscriptstyle
\text{H/KDEL}}$ to the cell periphery occurs within seconds 
and is immediately followed by the formation of plasma 
membrane-associated clusters within 20 minutes. Thereafter, 
cluster development follows an exponential growth and 
eventually saturates at time points $>80$ min. During this 
process, KDELR/cargo clusters are also internalized, however 
the precise temporal resolution of such endocytosis events 
has not yet been analyzed and will be subject of future studies.
Since all cell binding studies were performed under conditions 
of natural KDELR in vivo expression, and H/KDEL motifs on 
cellular proteins have solely been attributed to be exclusively 
recognized by KDELRs, our present data strongly point towards 
a function of KDELRs at the cell surface. Furthermore, our 
observation that ${\scriptstyle\text{eGFP-RTA}}^{\scriptscriptstyle
\text{H/KDEL}}$ shows a strong increase in toxicity and in vivo 
uptake (data not shown), likewise supports a role of KDELR-mediated 
cargo/toxin transport from the plasma membrane to the cytosol.

In an approach to characterize the observed clustering at the cell 
surface after cargo addition, we combined experiments and numerical 
methods and thereby demonstrate that cargo/KDELR cluster formation 
over time is a process that equally depends on temperature and cargo 
concentration. In particular, the later observation strongly points 
towards a regulated cellular mechanism to respond to an extracellular 
receptor ligand. Extensive simulations of cluster formation and size 
distribution indeed indicate that cells are capable to sense 
external cargo concentration and appropriately adapt the number 
of receptors at the plasma membrane. External KDEL-cargo addition 
likewise resulted in a response leading to an increase in anterograde 
receptor traffic to the plasma membrane. In addition, the lower 
cargo cluster numbers in the absence of active endocytosis and 
exocytosis (phalloidin-treatment or $4^{\circ}$C) indicate that under these 
conditions significantly less receptor molecules reach the plasma 
membrane, indirectly supporting our assumption that regulated 
KDELR transport is important for receptor cluster formation at 
the cell surface. One of the most striking features of the observed 
clustering is the power-law decay of cluster-size distribution 
P(s) with an approximately time-independent slope at longer times. 
To elucidate the origin of this behavior, we isolated and 
examined the role of surface dynamic processes such as 
receptor diffusion and receptor-receptor attraction in 
simulations, which mainly led to a fast (exponential-like) decay 
of the tail of P(s). We showed that preferential adsorption of 
receptors is a natural way to obtain adsorption kinetics and 
cluster-size distribution. Experimental data indicated that 
intracellular KDELR transport to the PM occurs along microtubules, 
and hot spots form in the vicinity of the regions where MTs approach 
the cell cortex and distribute anterograde arrival of KDELR-containing 
vesicles at the PM as well as collecting newly formed KDELR/cargo 
complexes from the cell surface by actin-mediated endocytosis. 
Both responses represent the dominant mechanisms that control 
receptor distribution at the cell surface, even if affected by 
additional factors such as e.g. receptor surface dynamics. We 
showed that both microtubule network and cortical actin are 
important key players in the clustering process, and inhibitation 
of MTs or filamentous actin strongly impaired the dynamic 
clustering at the cell surface. In future studies we will 
try to further dissect the underlying molecular processes 
and to identify the cellular components involved in KDELR/cargo 
cluster formation at the mammalian plasma membrane. In future 
studies we intend to use single-molecule cargo tracking in 
conjunction with high-resolution imaging to further dissect 
the underlying molecular processes and to identify the 
cellular components involved in KDELR/cargo cluster 
formation at the plasma membrane.

\subsection*{Methods}
\label{sec:Methods}
A detailed description of the experimental procedures and 
methods can be found in \emph{Supplementary Information}, 
including cultivation and transfection of mammalian cells, 
genetic techniques, affinity purification and immunochemical 
analysis of PM-localized KDELR1.  

\subsection*{Acknowledgments}
We thank K. Salo and L. Ruddock for the supply of cDNA of human KDELR1, 
L. Roberts for cDNA of RTA and anti-RTA antibody, and H. Rieger, K. 
Kruse and Z. Sadjadi for helpful discussions. This study was supported 
by the Deutsche Forschungsgemeinschaft (DFG) within the collaborative 
research center SFB 1027 (projects A6 and A7).

\subsection*{Author contributions statement}
B.B., M.J.S., M.R.S. and L.S. designed the research. D.R., T.B. and 
B.B. performed the experiments, M.R.S. analyzed the experimental 
data and performed simulations. M.R.S. and L.S. developed the 
theoretical framework and B.B., M.R.S. and M.J.S. wrote the 
manuscript. 
 
\subsection*{Competing financial interests} 
The authors declare no competing financial interests.

\end{document}